\documentstyle{article}
\begin{document}

\def\NPB#1#2#3{Nucl. Phys. B{#1} (19#2) #3}
\def\PLB#1#2#3{Phys. Lett. B{#1} (19#2) #3}
\def\PLBold#1#2#3{Phys. Lett. {#1B} (19#2) #3}
\def\PRD#1#2#3{Phys. Rev. D{#1} (19#2) #3}
\def\PRL#1#2#3{Phys. Rev. Lett. {#1} (19#2) #3}

\rightline{FTUAM-99/02; IFT-UAM/CSIC-99-02}
%\rightline{\tt hep-th/9804026}
\vspace{0.5cm}
\begin{center}
\LARGE{Recent Developments in Physics Far
Beyond the Standard Model
\footnote{Contribution to the  proceedings of the 
XXVI International Meeting on Fundamental Physics, La Toja, Spain (June 1998).} 
\\[20mm]}
%\small{(or why these theorists talk so much about dualities these days) \\[20mm]}
\large{L.E. Ib\'a\~nez \\[8mm]}
\small{
 Departamento de F\'{\i}sica Te\'orica C-XI
\\[-0.3em]
and \\[-0.3em]
Instituto de F\'{\i}sica Te\'orica  C-XVI,\\[-0.3em]
Universidad Aut\'onoma de Madrid\\[-0.3em]
Cantoblanco, 28049 Madrid, Spain
\\[7mm]}  
\small{\bf Abstract} \\[7mm]
\end{center}

\begin{center}
\begin{minipage}[h]{14.0cm}
String theory is the leading candidate for a unified
 theory of the standard model and gravity.
In the last few years theorists  have realized that there is a 
unique  structure underlying string theory. 
In this unification a prominent role is played by the 
duality symmetries which relate different theories.
I present a brief overview of these developements and 
discuss their possible impact in low-energy physics.
One of the lessons learned is that the string scale,
usually asumed to be of order of the Planck scale,
could be arbitrarily low, even close to 
accelerator energies.

\end{minipage}
\end{center}  
\newpage

\section{Far  Beyond the Standard Model (FBSM) }

You have already heard many times that the Standard Model (SM) is
fine and describes everything (except, maybe, neutrino masses)
very well. And you have also heard almost the same 
number of times that in spite of that there are plenty of aspects
of the SM that we do not yet understand. I am not going to repeat 
the well known list of questions without answers in the SM, it has not
changed very much in the last decade. Instead of that I am going
to talk today about one particular problem: the problem of 
combining the SM interactions with gravity. As is well known
naive quantization of Einstein's gravitation leads to a
non-renormalizable theory plagued with ultraviolate divergences.
 This problem has been with us 
for several decades and since fifteen years ago it is believed
that string theory is the best candidate for a solution. 
In addition we need a theory in which chiral gauge interactions
like those of the SM can coexist with quantum gravity. Also
this aspect seems to be present in strings  so let me   first of
all discuss  some key ingredients of these theories.

\section{String Theories}
The origin of ultraviolate divergences in perturbative gravity
(or any field theory for that matter) is related to the existence 
of interactions between particles which take place at  a point.
Insisting in pointlike interactions leads to ultraviolate divergences.
Whereas in the field theories of the $SU(3)\times SU(2)\times U(1)$
standard model those divergences can be absorved in the renormalization
process, that is not the case for gravity. String theories 
bypass this problem by avoiding point-like interactions and 
allowing the particles to interact in an extended region of space-time.

String theory \cite{string} 
is a radical departure from the standard understanding 
of physics because asumes that elementary (pointlike) particles
are not the fundamental blocks of nature. Rather the fundamental 
objects are strings, extended one-dimensional objects.
Whereas the movement of a classical particle is described by 
giving the dependence of coordinates on e.g., proper time $\tau $,
 $X^{\mu }(\tau )$ , the movement of a string in space-time is
described by giving $X^{\mu }(\tau , \sigma )$, where $\sigma $ 
is the coordinate along the string. Strings maybe either open 
(with two boundary points $X_a$, $X_b$ between which the string stretches)
or closed ($X_a=X_b$). Interactions of strings take place by joining 
and splitting. Thus e.g., a closed string propagating in space can
in a given moment split into two closed strings. In such a process of
splitting the interaction takes place in an extended region in the 
variables $(\tau , \sigma )$ and this is at the basis of the
absence of ultraviolate divergences in string theory.

An important property of string theory is that it predicts 
the existence of gravity. This comes about as follows. 
Let us consider the quantization of a closed string.
One can consider this system as a set of  an infinite 
number of harmonic oscilators with two type of vibration modes,
 right-handed (say) and left handed around the closed
string.
If  
 $a_n^{\mu }$  and ${\tilde a}_n^{\mu }$ ($n=0,\pm 1 , \pm 2 ...$) are
the two type of oscilators  one finds for the masses of the
string modes:
\begin{equation}
m^2\ = \ {1\over {\alpha '} }( N \ + {\tilde N} - 2 )
\label{masilla}
\end{equation}
where $N=\sum_{n} a_{-n}^{\mu }{a_{n}}^{\mu }$ and
 ${\tilde N}=\sum_{n} {\tilde a}_{-n}^{\mu }{{\tilde a}_{n}}^{\mu }$
are the corresponding number operators. Here ${1\over {\alpha '}}$ is the 
string tension and the $-2$ comes from the (regulated)  zero point 
vacuum energy. There is an additional constraint (coming from the non-existence 
of a privileged point on the string coordinate $\sigma $ ) which
imposses $N={\tilde N}$ for the eigenvalues of physical states.
Now, for $N={\tilde N}=1$ there is a {\it massless state with two vector
indices} , given by $g^{\mu \nu }=  a_1^{\mu }{\tilde a}_1^{\nu }|0>$.
Thus we have a massless spin two state in the spectrum, which is
no other but the graviton
\footnote{More specifically, it is the symmetrized state which gives
rise to the graviton. The antisymmetrized state gives rise to an 
additional antisymmetric field $A^{\mu \nu }$ and the trace
 to a massless scalar, the dilaton $\phi $ .}
.  Unfortunatelly for $N={\tilde N}=0$
we have also a tachyonic state with mass$^2$ equal to $-2/\alpha '$.
 This is signaling an instability in this (purely bosonic) string. 
Furthermore this theory has no fermions in the spectrum and hence
it cannot accomodate the observed quark and leptons.

Both problems are solved if we supersymmetrize the string variable
$X^{\mu }(\tau , \sigma )$ and include
fermionic coordinates $\psi ^{\mu }(\tau , \sigma )$. There are 
two types of closed superstrings of this type which go under the names
of Type IIA  and Type IIB \cite{type2} 
. They differ in the structure of the 
supersymmetry in both theories. The pairs $X^{\mu }, \psi ^{\mu }$
may be considered themselves as fields in a two-dimensional 
field theory on the variables $(\tau , \sigma )$, the world-sheet.
Both Type IIA and Type IIB have $N=2$ supersymmetry on the worldsheet
but the two SUSY generators have oposite chirality in the Type IIA case 
and the same chirality for Type IIB. In both cases one can show that,
in a explicitely supersymmetric formalism the tachyon disappears
from the massless spectrum. Furthermore they are finite theories
in perturbation theory without any divergence, neither ultraviolate 
nor infrared. Thus they seem to provide us for the first time with
consistent theories of quantum gravity.
However they have a puzzling property: they 
are formulated in ten  (one timelike and nine spacelike) dimensions.
Extra dimensions are not necesarily a problem since,
as already Kaluza and Klein showed in the twenties, they can be curled
up into a compact space at very short distances in such a way that
we only are able to see experimentally the standard four dimensions.
In the case of Type II strings, six dimensions are assumed to be 
compactified at very short distances $R$ in such a way that
only if we have energies higher than $1/R$ we would  be able to see
the six extra dimensions. Consistency with known gravitational
interactions suggest to identify :
\begin{equation}
1/R \ \propto {1\over {\sqrt{\alpha '}} } \ \propto M_{Planck} .
\label{escalas}
\end{equation}
The idea is that the SM particles (plus the graviton) correspond to
string modes which are (aproximately) massless compared to such huge
mass scales.

Even compactifying the six extra dimensions, Type II theories do not seem
very promising (at least in their pre-1995 formulation) to include
not only gravity but the standard model. We know that one of the most important
properties of the SM is its chirality: left-handed and right-handed
quarks and leptons transform differently under the SM group. Type IIA theory
is non-chiral and does not seem possible to include a chiral theory like
the SM inside it. Type IIB theory is on the other hand chiral but
it turns out that it does  not lead upon compactification to non-abelian
chiral theories like $SU(3)\times SU(2)\times U(1)$, it leads at most to
chiral $U(1)$ theories. 

Before 1984 only a third type of supersymmetric string,
also defined in ten dimensions, was known in
addition to Type II theories. It was the so called Type I theories.
This theory (unlike Type II) contain both closed and open strings. 
Closed (unoriented) strings give rise to the gravitational sector of the
theory and open strings give rise in general to non-Abelian gauge
theories. This is a very interesting property since one then has the 
hope of embedding  the gauge group $SU(3)\times SU(2)\times U(1)$
into the open string sector. Unfortunatelly it was realized 
\cite{anom}
that the theory has  anomalies and hence it is
inconsistent at the quantum level. Thus before 1984 the situation
concerning string theory was very puzzling: Type II theories 
had a consistent quantum theory without anomalies but where 
unable to embedd the SM. On the other hand Type I strings
had the potential to embed the SM but were anomalous!
One thus can understand why before 1984 only  very, very
few theorists were working in the field of string theory.

In 1984 everything changed when it was realized that 
the conclusion that  Type I string is  inconsistent was
in fact not true. Green and Schwarz showed \cite{gs} 
 that for a very particular
gauge group, $SO(32)$,  all anomalies cancelled
via a new and ellegant mechanism. Soon after
Gross, Harvey, Martinec and Rohm
\cite{heter}
 showed the existence of two
more string theories in ten dimensions, the heterotic strings.
They are theories that, like Type I,
are chiral and  contain gauge fields in
their massless spectra. However they only contain closed 
strings. The gauge groups are either $SO(32)$ or $E_8\times E_8$
and are also anomaly free.
These  developements caused the well known increase of 
popularity of this field. Now we had three string theories
which were candidates to unify gravitation and the SM into
a finite theory. Of course, being ten-dimensional
a process of compactification down to the four physical
dimensions is  required. Most phenomenological attempts
to embed the SM were based in the heterotic $E_8\times E_8$
string since, as shown by Candelas, Horowitz, Strominger and Witten
\cite{phil} ,
there are particular classes of compact six-dimensional 
varieties (known as Calabi-Yau manifolds) such that,
upon compactification of the $E_8\times E_8$ string 
give rise at low energies to an $E_6$ gauge theory 
with one $N=1$ supersymmetry (usefull to solve the hierarchy 
problem) and 
a number of fermion generations given by 
one half the Euler characteristic of the Calabi-Yau 
manifold. The gauge group $E_6$ had been used as a grand
unification group in the past so everything looked like if
we were really close to the final unified theory of the
standard model and gravity. 

This optimism was cooled down when people tried to get a
more specific contact with low energy phenomenology. 
It was soon realized that although $E_6$ is a nice group, is
still far away from what we really want : the SM with three
quark-lepton generations, apropriate Higgs doublets,
a reasonable fermion mass spectrum, sufficiently stable proton, etc.
A  number of techniques (orbifolds, fermionic construction
etc.) were developed to get miriads of new possibilities for
heterotic string compactifications leading at low energies to 
plenty of possibilities for gauge group, fermion content and
phenomenology. A few  of the new string vacua obtained with the
new techniques had massless spectrum quite close to that of the minimal 
supersymmetric SM (MSSM), which is already by itself a remarkable
achievement \cite{quev}
. In spite of this success, the presence of such miriads 
of aparently consistent string vacua leads to an obvious question:
how is the physical vacuum (presumably corresponding to the
SM ) chosen by the dynamics of the theory? With unbroken
supersymmetry all string vacua are degenerate,  which leads us to the
next  question: how is supersymmetry broken in string theory ?
Both questions seem to require a knowledge of non-perturbative
effects in string theory.

\section{T-duality}

By 1985 the five existing supersymmetric $D=10$ string theories
had already been discovered: Type IIA, Type IIB, $SO(32)$ Type I,
$SO(32)$ heterotic and $E_8\times E_8$ heterotic. As we said,
much  effort was dedicated to the study of the 
heterotic strings and also to the study of two-dimensional 
conformal field theory (CFT) in general, the latter being relevant to
study the properties of the string variables $X(\tau , \sigma )$
in the two-dimensional world-sheet spanned by $\tau $ and $\sigma $.
Soon a peculiar symmetry of closed strings was noticed, which
now goes under the name of T-duality. Consider now for simplicity 
the case of the purely bosonic string with coordinates
$X^{\mu }(\tau , \sigma )$ which we discussed at the beginning of
previous section. The simplest example of compactification is one in which
only one dimension is curled up forming a microscopic circle of
radius $R$. One can see that the mass$^2$ of the string modes has
now the form \cite{string} :
\begin{equation}
m^2\ = \  ( {{p^2}\over {R^2} }\ +\  { {w^2R^2}\over { {\alpha '}^2}  })\ +\ 
{1\over {\alpha '} }( N \ + {\tilde N} - 2 )
\label{tdual}
\end{equation}
where $p,w=0,\pm 1,\pm 2,...$. In this formula the first term in the
rhs. is not 
special of string theories, it just corresponds to the fact that 
in quantum mechanics with one dimension of finite size, there are
quantized momemtum ($p$) states, the
Kaluza-Klein modes in our case. The second term in the rhs. 
(proportional to $R^2$) is purely stringy, as is obvious from the fact that
$\alpha '$ appears.  It represents the possibility that a closed string
winds  $w$ times around the circle of radius $R$. The above formula
has an interesting property observed \cite{ky} by K. Kikkawa and M. Yamasaki in 
1984. It remains invariant under the replacement:
\begin{equation}
p\leftrightarrow w  \ \ ;\ \ R \leftrightarrow
{ { {\alpha '} }\over {R} } ,
\label{recambio}
\end{equation}
i.e., we exchange quantized momenta $p$ by winding number $w$
and  at the same time $R$ by $\alpha '/R$. Physically this is quite a
surprising symmetry since it indicates that the energies of 
string states are identical in a theory with compact radius  $R$
and in a theory with radius $\alpha '/R$. It turns out that not only the
spectrum but all physical properties of those two string configurations
are identical. This is remarkable because this tells us that 
for a compactified string large $R$ and small $R$ are equivalent!
This is the simplest example of what now is called T-duality
\footnote{This symmetry was termed just duality.
It was first called T-duality in ref.\cite{filq} 
 to distinguish it from
the newly proposed S-duality.}.
If instead of one dimension of radius $R$ one compactifies two
dimensions, the $R\leftrightarrow \alpha '/R$
duality generalizes in an interesting way. The apropriate 
duality transformation is now \cite{shapwil} :
\begin{equation}
T\ \rightarrow \ {1\over T} \ \ \ ;\ \ \ T\ \rightarrow T+1 
\label{tdual2}
\end{equation}
where now $T$ is a complex ("modulus") field  $T=b +iR^2$. Here
$R^2$ is  the overall compact volume and $b$ is an axion-like
field. The above two transformations generate the discrete
infinite group $SL(2, Z)$. Again, this transformation has to be
acompanied by the apropriate transformation of quantized 
momenta and winding numbers.

When applied to the list of ten-dimensional closed supersymmetric strings
a number of equivalences were found. It turns out that
Type IIA string compactified on a circle of radius $R$ is equivalent 
(T-dual) to Type IIB compactified on a circle of radius $\alpha '/R$
(and viceversa) \cite{dhs} .
Furthermore the heterotic $E_8\times E_8$ compactified on
a circle of radius $R$ is T-dual to 
the $SO(32)$ heterotic compactified on a circle of radius $\alpha '/R$
(and viceversa) \cite{pablito}
. This is remarkable since it shows us that there are 
indeed only three types of supersymmetric strings which are
disconnected: Type II, Type I and heterotic. Thus the number of candidates
for a fundamental theory was substantially reduced. This nevertheless did
not impress very much the string practitioners. Tacitally most stringers 
had in the back of their minds the idea that only the heterotic
string had any chance of constituting the long sought for
unified theory of all known interactions. Type II and Type I strings
were bothersome theories that sooner or later would be proven to be 
inconsistent. In this way we would be left with a unique unified
theory, the heterotic. This turned out not to be the case, as 
has been shown in the last four years.

\section{S-duality}

In string theory the strength of interactions is not given by 
coupling constants but by the vacuum expectation value of
a massless real scalar $\phi $, the dilaton. It turns out that, although
apparently totally different in origin, the dilaton behaves 
to some extent in a way quite similar to the compact radius $R$
which appears in the field $T$ described in the previous section.
Let us consider the simplest compactification of the heterotic string
down to four dimensions obtained by compactifying on a six torus.
The low energy Lagrangian has $N=4$ supersymmetry in $D=4$.
It turns out that the dilaton appears naturally as the imaginary
part of a complex scalar field $S=\eta + i \phi $.
Here $\eta $ is an axion-like field. + Furthermore,
at the classical level the field $S$ appears in the Lagrangian in
a way quite analogous to the fields $T$. Both have asociated
symmetries of the type $SL(2,R)$. We already mentioned above that
in the case of the $T$ field a discrete subgroup of that symmetry, 
  $SL(2,Z)$,  was in fact exact,
corresponding to $T$-duality. A natural conjecture would then be
to assume that, also for the $S$ field an $SL(2,Z)$ symmetry
generated by the transformations:
\begin{equation}
S\ \rightarrow \ {1\over S} \ \ \ ;\ \ \ S\ \rightarrow S+1 
\label{sdual}
\end{equation}
could be a symmetry of this class of string vacua
and this was our proposal \cite{filq}
in 1990. But now this symmetry
is quite bizarre: notice that eq.(\ref{sdual}) includes the
transformation $\phi \leftrightarrow 1/\phi $. Since the dilaton 
$\phi $ gives a messure of the strength of the interactions, this is
a symmetry which relates strongly coupled to weakly coupled string
theory! Unlike the case of $T$-duality this symmetry is clearly
non-perturbative in nature and hence difficult to check
with purely perturbative methods.

The origin of the $SL(2,Z)$ symmetry in the case of  T-duality 
had to do with the generalization of the simple symmetry
$R\leftrightarrow \alpha '/R$ which appears when a closed string has
one compact dimension of size $R$. Is there anything like this
in S-duality? In fact it turns out that if one derives
the Lagrangian of $D=10$ supergravity starting from $D=11$
supergravity and compactifies  on
a circle of radius $R_{11}$, the ten-dimensional dilaton $\phi = R_{11}$,
and hence one would have hoped 
that perhaps a duality in $R_{11}$
would have then implied a $\phi \leftrightarrow 1/\phi $
symmetry  \cite{filq} .
But in 1990 $D=11$ supergravity had no room in the realm of
string theory and we had to wait till 1995 to have a geometrical
understanding of  S-duality. 

In the case of T-duality, the transformations of eq.(\ref{tdual2})
come along with the interchange of quantized momenta $p$ 
and winding numbers $w$. What is the equivalent statement in
S-duality? We proposed \cite{filq} that in the case of S-duality
one has to exchange standard charged particles with
 solitonic monopole states which generically are present in 
this class of theories. In fact already in 1975 
Claus Montonen and David Olive \cite{montolive}
had considered the posibility 
of an exchange symmetry between charged particles and monopoles
in certain non-supersymmetric  gauged field theories with scalars.
They  conjectured the possibility that the physics of charged
particles at strong coupling $g^2$ could be equivalent to that of
magnetic monopoles at small coupling. In the $N=4$ supersymmetric
version of these theories \cite{susymo} 
one finds  certain states, called 
BPS states whose masses are given by an expression of the form:
\begin{equation}
M^2\ \propto \ q_e^2g^2 \ +\ { {q_m^2}\over {g^2} }
\label{molive}
\end{equation}
where $q_e$($q_m$) are the electric(magnetic) charges of the particles
and $g^2$ is the gauge coupling constant$^2$. Notice the explicit 
invariance under the exchanges  $q_e\leftrightarrow q_m$ ,
$g^2\leftrightarrow 1/g^2$.
It is obvious the analogy
of this expression 
to that of the first two terms in the rhs. of eq.(\ref{tdual}) .
In the string case one would have the dilaton $\phi $ instead of $g^2$.
Since in the heterotic string compactification that we are discussing
there is also an effective $N=4$ supersymmetric Lagrangian,
 we considered this as additional circumstantial evidence in favour
of the S-duality symmetry \cite{filq} . 

Being a non-perturbative symmetry, it was not obvious how to 
find evidence in favour of its reality. However A. Sen
\cite{asen} 
studied in more detail the heterotic string compactified on the
six torus and found that the spectrum of BPS particles was
indeed invariant under S-duality. More importantly, starting with a
purely electric state,  S-duality predicts the existence of 
certain monopole configurations obtained by $SL(2,Z)$ transformations.
Ashoke Sen found those solutions with the correct multiplicities
\cite{asen}.
His  arguments applied even non-perturabatively because the mass-formulae
for BPS states is necesarily exact due to $N=4$ supersymmetry.
These developements changed the initially skeptical atitude
of the comunity concerning strong-weak coupling dualities and gave
rise to the beginning of the so called second  revolution
of string theory \cite{sreviews} 
. It also inspired the seminal work of
Seiberg and Witten \cite{sw}
in which they constructed the exact 
effective Lagrangian of certain gauge field theories
with $N=2$ supersymmetry 
in four dimensions by using duality arguments.

\section{Dualities and unification of all string theories}

The S-duality symmetry of the heterotic string compactified on a
six-torus which we discussed in the previous section was the first 
example of a general class of symmetries present in many
string configurations. Starting in 1994 many such strong/weak coupling
dualities were conjectured to hold. These conjectures have not been
formally proven but a number of consistency checks have been made 
in a good number of cases making difficult to believe that they are
purely accidental. It is dificult to briefly describe all the concepts about
string theory that have changed in the last four years
due to these dualities. We will
limit ourselves here to review some of the results which are more 
directly related to the unification of all string theories into a single
structure. In this context the pioneering work of M. Duff \cite{duff} ,
A. Strominger \cite{strom} ,  P. Townsend and 
C. Hull \cite{paul} , completed and systemathized by E. Witten in 1995
\cite{witt}  were very
important.

We already  discussed how, due to T-duality, there are only three disconnected
ten-dimensional supersymmetric string theories: Type I, Type II and heterotic.
The new S-dualities further reduce the number  of independent 
theories.
It has been found that \cite{sreviews} :

 \begin{enumerate}

\item
Type I string theory and the heterotic $SO(32)$ are S-dual to each other
\cite{polwit} .
This means that ten-dimensional
Type I theory at small coupling $\phi $ is equivalent to
heterotic $SO(32)$ at  (large) $1/\phi $ coupling. This is in some way
the least surprising of all dualities since both theories posses 
$SO(32)$ gauge bosons.

\item

Type IIB theory is S-dual to itself \cite{paul} .  

\item

Type IIA theory gives rise to a surprise \cite{witt}
. At weak coupling $\phi $ it is
a ten-dimensional theory but increasing the coupling a new eleventh 
dimension reveals itself. Type IIA theory corresponds to a new theory
in eleven dimensions in which one of the dimensions
is compactified on a circle with radius $R_{11}=\phi $. The Kaluza-Klein
states in this 11-th dimension have masses $\propto 1/\phi $ and that 
is why at weak coupling (small $\phi $ )  this extra dimension is not
seen. This is a new surprising phenomenon:
as the strength of interactions increases new dimensions may appear.
This 11-dimensional theory, whose low-energy Lagrangian turns out to coincide
with that of 11-dimensional supergravity, is termed  M-theory and 
a proper formulation for it is still lacking.

\item

Something analogous to the Type IIA case happens for the 
$E_8\times E_8$ heterotic \cite{horwit}
. As one increases the coupling $\phi $ a new
eleventh dimension reveals itself with size $R_{11}=\phi $.
However in this case the $E_8\times E_8$ theory is obtained 
by compactifying the 11-dimensional M-theory on a finite segment
of length $R_{11}$. At each of the two boundary points of the
segment one gets a gauge group $E_8$ and this explains the 
presence of the two factors.

\end{enumerate}

With all the above connections and equivalences it is clear that
all known string theories are connected to each other and to the
misterious 11-th dimensional M-theory: Type IIA and $E_8\times E_8$
theories are both connected to M-theory at strong coupling. On the other hand
T-dualities  connect those two theories with Type IIB and $SO(32)$
heterotic respectively. Finally, this last theory is connected by
S-duality to Type I $SO(32)$ theory. Thus all five supersymmetric string
theories are in one way or another connected to M-theory.
All string theories are unified.

\section{Dualities and p-branes}

An important role in the duality developements is played by
p-branes \cite{sreviews,poldb} 
 . They are extended objects which are generalizations
of the notion of string. We mentioned how the string coordinate
$X^{\mu }(\tau , \sigma )$ depends on one time-like variable 
$\tau $ and a space-like variable $\sigma $ which parametrizes
the position along the string. As the string moves it sweps
a world-sheet parametrized by $(\tau , \sigma )$.
 One can as well consider other
extended objects  (p-branes) 
whose coordinates $X^{\mu }(\tau, \sigma _a )$,
$a=1,..,p$ depend on more than one space-like variable $\sigma _a$.
As they move they swep a $p+1$-dimensional "world-volume".
Point-like particles are 0-branes,  strings are  1-branes,
membranes are 2-branes etc. 
In fact these p-branes appear as soliton-like solutions in
the effective low-energy field theories from the different 
supersymmetric strings. This is very much analogous to 
monopoles,  which also appear as soliton-like solutions of
certain gauge theories with scalar fields.
In particular, the massless sector of supersymmetric 
strings contain antisymmetric tensor fields
$A_{\mu _1 ..\mu _n}$. It turns out that one finds 
p-brane solutions for $p=n-1$ ("electric") and for
$p=D-n-3$ ("magnetic"), where $D$ is the number of
space-time dimensions. Thus, for example, in all
$D=10$ closed  strings there is a  massless
antisymmetric tensor field $B_{\mu \nu }$.
Thus $n=2$ and we have a 1-brane (the fundamental 
 string)  and a 5-brane.

In the case of Type II strings in addition one has some
extra  massless antisymmetric fields with odd(even)  number
of indices for Type IIA(B). Thus there are p-branes 
with even(odd) p  for TypeIIA(B) strings,
with $p<10$. These are called Dirichlet branes \cite{poldb} , 
D-branes for short,  and turn out to be particularly 
important because they allow for some specific 
perturbative computations which lead to checks of
certain duality conjectures.
One important property of these D-branes is that 
open Type I strings are forced to have their 
boundary ends lying necesarily on the portion of space
ocupied by D-branes. In the case of 9-branes
their worldvolume ocupies all ten dimensions and hence
the end-points of an open string may be anywhere.
But for $p<9$ open strings are forced to start and
end  on  submanifolds of the full 10-dimensional 
space, the submanifold ocupied for the
relevant D-brane worldvolume. Thus, for example, in the case of 3-branes
whose world-volume can be made to coincide with the
physical Minkowski space, open strings are forced to
start and end on Minkowski space, they cannot 
start or end on the remaining $10-4=6$ dimensions
(the famous "bulk").

\section{Embeding the observed world into D-branes}

D-branes have asociated gauge fields living in their
worldvolume \cite{poldb} 
. Due  to this fact, open strings starting and ending
on the same D-brane give rise to a massless $U(1)$ field. If we have
$N$ D-branes with overlapping worldvolume the gauge symmetry is
enhanced to $U(N)$. The non-Abelian generators correspond to
open strings going from one D-brane to another. A variety of
Type II or Type I string vacua  can be constructed
containing different p-brane configurations considered as
static classical objects. This new class of string vacua
may have in general quite large gauge groups, groups
with rank much higher than 28  which is the maximum for
a $D=4$ heterotic compactification. This already shows us
that the possilities for embedding the physical
$SU(3)\times SU(2)\times U(1)$ group into string theory  
 are much wider than in the perturbative heterotic
vacua in which the SM group was always a subgroup
either of $E_8\times E_8$ or $SO(32)$. Thus in principle one can
now consider the SM group as coming from a configuration
of e.g., four coinciding 3-branes. Although this is true
in principle, getting chirality and 3 generations
in this manner  is not
so obvious at the moment.
New Type I $D=4$, $N=1$ string models have recently been built
\cite{fdorient} 
with the new techniques which incorporate these new features
although it is too soon to expect fully realistic models.

If indeed the SM gauge group corresponds to gauge fields
living on the world-volume of 3-branes
\footnote{Or (q+3)-branes with q dimensions wrapping on the
compact dimensions.} some surprises concerning the structure
of fundamental mass scales appear
 \cite{uniwitten,lykken,untev,anton,otros,biq,imr}
.  Let me first recall the
relationship between Planck mass $M_{Planck}$ and string 
scale $M_s=1/\sqrt{\alpha ' }$. In the perturbative heterotic
string those two scales are unavoidably tied up by the equation
$M_s^2=\alpha _{GUT} M_{Planck}^2/8$, where $\alpha _{GUT}$
is the gauge coupling. This is in some way related to the fact
that both gravitational and gauge interactions come from closed
strings, there is a real unification  between both
interactions. Thus the string scale is necesarily closed to the
Planck mass, $M_s \simeq 10^{17}$ GeV.
An important point is that this relationship between  $M_s$
and $M_{Planck}$ is  (in first aproximation) 
independent of the compactification 
scale $M_c$ which gives as the overall inverse size of the 
six extra compact dimensions.

This is not in general the case for Type I string vacua. In this case, if
the gauge group comes from open strings starting and ending on a set
of p-branes,  one has \cite{uniwitten,anton,otros,imr} :
\begin{equation}
{ {M_c^{(p-6)}} \over {M_s^{(p-7)} } } \ =\ { {\alpha _p M_{Planck} }\over {\sqrt {2} }  }
 \
\label{const}
\end{equation}
where $\alpha _p$ is the gauge coupling of the corresponding gauge group.
Consider for example the case in which we asume that our gauge
group comes from a set of 3-branes ($p=3$). Then one has
\begin{equation}
M_s^4 \ =\ { { \alpha _{GUT}} \over {\sqrt{2} }}
 M_c^3M_{Planck}  \ \ .
 \label{3branas}
\end{equation}
It is clear from this expression that one can lower the value of the
string scale $M_s$ as much as we wish by lowering accordingly
the compactification scale $M_c$ and still mantaining $M_{Planck}$
fixed at its experimental value. Thus the standard statement that
string physics happens  at the Planck scale is not justified in the
present situation. The value of $M_s$ depends on what the size
of the extra dimensions are. 

One could argue that the compactification scale $M_c$ cannot be
below a few hundred GeV, since accelarators have not seen
any Kaluza-Klein excitations at those energies. However this conclusion
would have been incorrect, as recently emphasized in ref.\cite{untev} . 
Indeed, I already mentioned above that 
the gauge group coming from 3-branes lives only
on Minkowski space since open strings can only
start or end on the worldvolume of the 3-branes,
which we have identified with Minkowski space.
Open strings cannot propagate in the bulk. 
Other way of saying the same thing is that 
the gauge group has no Kaluza-Klein excitations.
On the other hand the gravitational sector, 
which comes from closed strings, can propagate 
in the bulk and has both Kaluza-Klein and winding modes.
However this Kaluza-Klein gravitational excitations 
are very weakly coupled and would not have  been observed
at accelerators even if 
$M_c$ was as small as the weak scale. 

What is then the scale of string theory $M_s$?
It cannot be  lower than  1 TeV or so
since otherwise string excitations would have been
observed at accelerators. Notice that although
the gauge group has no KK excitations it has string excitations.
Thus in principle any value of $M_s$ above 1 TeV is possible.
There are however three natural  options for the value of $M_s$
which look particularly appealing:

{\bf i)} $M_s\simeq M_{GUT}$ \cite{uniwitten}
 . Here $M_{GUT}$ is the scale
at which the extrapolated gauge couplings of the minimal
supersymmetric standard model (MSSM) join. Numerically
this is of order $10^{16}$ GeV. This is compatible with
eq.(\ref{3branas}) for values of $M_c$ slightly 
below $M_s$. This possibility has the advantage that
it incorporates the succesfull joining of coupling 
constants of the MSSM in a natural way. On the other hand
the gauge hierarchy between $M_W$ and $M_{Planck}$ has to be
blamed to some mechanism (like gaugino condensation) able
to generate such large hierarchies.

{\bf ii)} $M_s\simeq \sqrt{M_WM_{Planck} }$ . This is the
geometrical intermediate scale $\simeq 10^{11}$ GeV which 
coincides with the SUSY-breaking scale in models with a
hidden sector and gravity mediated SUSY breaking.
The interest of this choice has been recently emphasized in
ref.\cite{biq}  . In this case one has $M_c/M_s\simeq \alpha \simeq 0.01$
and the hierarchy between $M_W$ and $M_{Planck}$ may be understood
without the necessity of any hierarchy-generating mechanism
like gaugino condensation. In addition in this scheme it
is easy to accomodate the axion solution to the strong CP problem.
Indeed, cosmological and astrophysical constraints imply that
the axion scale should be of order of the intermediate scale.
On the other hand axion fields with this scale appear naturally
in Type I string models with $M_s$ of this order.

{\bf iii)} $M_s\simeq $ 1 TeV \cite{lykken,untev,anton}
.  In this case the hierarchy 
problem is solved in an obvious way by lowering the
fundamental scale of gravitation close to the weak scale.
In this case eq.(\ref{3branas}) gives
 (for an isotropical compactification)
$M_c/M_s\simeq 10^{-5}$  and hence $M_c\simeq $ 10 MeV.
As we remarked above, such low values of $M_c$ are perfectly
compatible with accelerator data since gauge fields have no KK
excitations. In fact one can even consider a non-isotropic
case with four compact  dimensions with $M_c\simeq $ 1 TeV and
the other two compact dimensions with $M_c\simeq 10^{-3}$ eV
\cite{untev,anton} .
In this case string physics could be tested at future accelerators
such as LHC. On the other hand the large  hierarchy between 
$M_c$ and $M_s$ has now to be explained.

Each of these three possibilities has its own adventages and shortcomings,
the first being the most conservative one. The possibility of finding 
traces of string theory at accelerator energies as may happen
in the 1 TeV scenario is quite exciting
\cite{bursta} . On the other hand a number
of theoretical issues like proton stability, gauge coupling unification,
cosmology and generation of the $M_c/M_s$ hierachy may be difficult to deal with
in that scheme. All those questions seem more easy to deal with in the
scheme with $M_s$ equal to the intermediate scale. 
A lot of effort is at present being dedicated to the study of the different
alternatives.

\section{Epilogue}

The duality revolution has changed not only our ideas
about string theory but also about field theory. 
Many perturbative and non-perturbative properties of
supersymmetric gauge theories have been reformulated in the last
three years in the language of D-branes giving a geometrical
interpretation to facts like the Higgs effect or
Seiberg's duality. There are reasons to believe that the
D-brane language may lead eventually to a more fundamental 
formulation of gauge theories themselves.
A new duality has also been propossed by
Maldacena  \cite{malda} 
which relates certain gauge field theories
to Type II string theories on certain backgrounds.
That has lead even to attempts \cite{glue} 
 to compute
the glueball spectrum of QCD in terms of
the dual Type IIB compactified theory. Although
it is not clear whether the aproximations made
in such computations are in a consistent regime,
the mere fact of thinking that one can perhaps 
compute physical things like glueball masses 
show how string theory and duality are giving us
unexpected tools for the study of field theories.

An important application of D-branes has been to the 
 problem of information loss in blackholes.
In particular, for certain classes of blackholes,
it has been possible \cite{sv}  to obtain a microscopic 
explanation of the Bekenstein-Hawking formula in
terms of D-branes (see e.g. refs.\cite{bek}  for an intuitive explanation of
this and references).
On the other hand we are still lacking a fundamental
formulation of  M-theory, the theory which unifies
all supersymmetric string theories and 11-dimensional
supergravity. There is a candidate  for
such a fundamental non-perturbative formulation of M-theory 
which goes under the name of  M(atrix)-theory \cite{matrix} 
 (see refs.\cite{bek}
for an intuitive idea of this). 
A lot of work remains to be done in order to understand
what is the fundamental theory which incorporates 
the notion of dualities in a built-in manner. It is sure
that many surprises are still to come both from the
more theoretical
as well as from the more phenomenological sides.

\bigskip

\bigskip

\leftline {\bf Acknowledgments}
I thank the organizers of this meeting for their kind invitation
to this wonderfull place. And I apologize to those members of the
audience  who got tangled up in the
network of dimensions, dualities and compactifications  they were exposed to
during my talk.

\end{document}